\documentstyle[12pt,epsfig,amss,wrapfig]{article}

\catcode`\@=11
\long\def\@makefntext#1{
\protect\noindent \hbox to 3.2pt {\hskip-.9pt
$^{{\ninerm\@thefnmark}}$\hfil}#1\hfill}                

\def\@makefnmark{\hbox to 0pt{$^{\@thefnmark}$\hss}}  

\def\ps@myheadings{\let\@mkboth\@gobbletwo
\def\@oddhead{\hbox{}
\rightmark\hfil\ninerm\thepage}
\def\@oddfoot{}\def\@evenhead{\ninerm\thepage\hfil
\leftmark\hbox{}}\def\@evenfoot{}
\def\sectionmark##1{}\def\subsectionmark##1{}}

\renewcommand{\thefootnote}{\fnsymbol{footnote}}

\newcounter{sectionc}\newcounter{subsectionc}\newcounter{subsubsectionc}
\renewcommand{\section}[1] {\vspace*{0.6cm}\addtocounter{sectionc}{1}
\setcounter{subsectionc}{0}\setcounter{subsubsectionc}{0}\noindent
        {\normalsize\bf\thesectionc. #1}\par\vspace*{0.4cm}}
\renewcommand{\subsection}[1] {\vspace*{0.6cm}\addtocounter{subsectionc}{1}
        \setcounter{subsubsectionc}{0}\noindent
        {\normalsize\it\thesectionc.\thesubsectionc. #1}\par\vspace*{0.4cm}}
\renewcommand{\subsubsection}[1]
{\vspace*{0.6cm}\addtocounter{subsubsectionc}{1}
        \noindent {\normalsize\rm\thesectionc.\thesubsectionc.\thesubsubsectionc.
        #1}\par\vspace*{0.4cm}}

\newcounter{appendixc}
\newcounter{subappendixc}[appendixc]
\newcounter{subsubappendixc}[subappendixc]

\renewcommand{\appendix}[1] {\vspace*{0.6cm}
        \refstepcounter{appendixc}
        \setcounter{figure}{0}
        \setcounter{table}{0}
        \setcounter{equation}{0}
        \renewcommand{\thefigure}{\Alph{appendixc}.\arabic{figure}}
        \renewcommand{\thetable}{\Alph{appendixc}.\arabic{table}}
        \renewcommand{\theappendixc}{\Alph{appendixc}}
        \renewcommand{\theequation}{\Alph{appendixc}.\arabic{equation}}
        \noindent{\bf Appendix \theappendixc #1}\par\vspace*{0.4cm}}

\def\abstracts#1{{
        \centering{\begin{minipage}{12.2truecm}\footnotesize\baselineskip=12pt\noindent
        \centerline{\footnotesize ABSTRACT}\vspace*{0.3cm}
        \parindent=0pt #1
        \end{minipage}}\par}}

\renewenvironment{thebibliography}[1]
        {\begin{list}{\arabic{enumi}.}
        {\usecounter{enumi}\setlength{\parsep}{0pt}
\setlength{\leftmargin 1.25cm}{\rightmargin 0pt}
         \setlength{\itemsep}{0pt} \settowidth
        {\labelwidth}{#1.}\sloppy}}{\end{list}}

\newcounter{itemlistc}
\newcounter{romanlistc}
\newcounter{alphlistc}
\newcounter{arabiclistc}

\newcommand{\fcaption}[1]{
        \refstepcounter{figure}
        \setbox\@tempboxa = \hbox{\footnotesize Fig.~\thefigure. #1}
        \ifdim \wd\@tempboxa > 6in
           {\begin{center}
        \parbox{6in}{\footnotesize\baselineskip=12pt Fig.~\thefigure. #1}
            \end{center}}
        \else
             {\begin{center}
             {\footnotesize Fig.~\thefigure. #1}
              \end{center}}
        \fi}

\newcommand{\bild} [2]{
                       \refstepcounter{figure}
                       {\begin{center}
     \parbox{#1}{\footnotesize\baselineskip=12pt Fig.~\thefigure. #2}
                       \end{center}}
                       }
\newcommand{\tcaption}[1]{
        \refstepcounter{table}
        \setbox\@tempboxa = \hbox{\footnotesize Table~\thetable. #1}
        \ifdim \wd\@tempboxa > 6in
           {\begin{center}
        \parbox{6in}{\footnotesize\baselineskip=12pt Table~\thetable. #1}
            \end{center}}
        \else
             {\begin{center}
             {\footnotesize Table~\thetable. #1}
              \end{center}}
        \fi}

\def\@citex[#1]#2{\if@filesw\immediate\write\@auxout
        {\string\citation{#2}}\fi
\def\@citea{}\@cite{\@for\@citeb:=#2\do
        {\@citea\def\@citea{,}\@ifundefined
        {b@\@citeb}{{\bf ?}\@warning
        {Citation `\@citeb' on page \thepage \space undefined}}
        {\csname b@\@citeb\endcsname}}}{#1}}

\newif\if@cghi
\def\cite{\@cghitrue\@ifnextchar [{\@tempswatrue
        \@citex}{\@tempswafalse\@citex[]}}
\def\citelow{\@cghifalse\@ifnextchar [{\@tempswatrue
        \@citex}{\@tempswafalse\@citex[]}}
\def\@cite#1#2{{$\null^{#1}$\if@tempswa\typeout
        {IJCGA warning: optional citation argument
        ignored: `#2'} \fi}}

 1
 1
 1

\font\ninerm=cmr9

\newcommand{\av}[1]{\mbox{$ \langle #1 \rangle $}}
\newcommand{\ol}[1]{\mbox{$\overline{#1}$}}
\newcommand{\order}[1]{\mbox{${\cal O}(#1)$}}
\newcommand{\ncs}{\mbox{$N_{\rm CS}~$}}
\newcommand{\ein}{\mbox{$E_{\rm in}~$}}
\newcommand{\eout}{\mbox{$E_{\rm out}~$}}
\newcommand{\einp}{\mbox{$E_{\rm in}'~$}}
\newcommand{\eoutp}{\mbox{$E_{\rm out}'~$}}

\newcommand{\qprimesq}{\mbox{$Q'^2~$}}
\newcommand{\xprime}{\mbox{$x'~$}}

\newcommand{\qprimex}{\mbox{$Q'$}}
\newcommand{\qprimesqx}{\mbox{$Q'^2$}}
\newcommand{\xprimex}{\mbox{$x'$}}

\newcommand{\flim}{\mbox{$f_{\rm lim}~$}}
\newcommand{\finst}{\mbox{$f_{I}~$}}

\newcommand{\shat}{\mbox{$\hat{s}$}}

\newcommand{\W}{\mbox{$W~$}}

\newcommand{\xb}{\mbox{$x~$}}  

\newcommand{\Qsq}{\mbox{$Q^2~$}}

\newcommand{\et}{\mbox{$E_T~$}}

\newcommand{\pt}{\mbox{$p_T~$}}

\newcommand{\nmax}{\mbox{$n_{\rm max}~$}}

\newcommand{\dd}{{\rm d}}

\newcommand{\TeV}{\mbox{\rm ~TeV~}}
\newcommand{\TeVx}{\mbox{\rm TeV}}
\newcommand{\GeV}{\mbox{\rm ~GeV~}}
\newcommand{\GeVx}{\rm GeV}

\newcommand{\GeVsq}{\mbox{${\rm ~GeV}^2~$}}
\newcommand{\GeVsqx}{\mbox{${\rm GeV}^2$}}

\newcommand{\pbx}{\mbox{${\rm pb}$}}

\newcommand{\pbinv}{\mbox{${\rm ~pb^{-1}}~$}}

\begin{document}

\noindent
{\tt DESY 97-165    \hfill    ISSN 0418-9833} \\
{\tt MPI-PhE/97-21} \\
{\tt hep-ph/9709206} \\
{\tt August 1997}                  \\

\centerline{\normalsize\bf
QCD INSTANTONS AT HERA\footnote{invited talk at the
Ringberg Workshop ``New Trends in HERA Physics'',
Schlo\ss\ Ringberg, Tegernsee, May 1997.}}
\baselineskip=22pt


\centerline{\footnotesize MICHAEL KUHLEN}
\baselineskip=13pt
\centerline{\footnotesize\it Max-Planck-Institut f\"ur Physik,
Werner-Heisenberg-Institut}
\baselineskip=12pt
\centerline{\footnotesize\it F\"ohringer Ring 6, D-80805 M\"unchen, Germany}
\centerline{\footnotesize E-mail: kuhlen@desy.de}

\vspace*{0.9cm}
\abstracts{
Phenomenological aspects of chirality violating processes
induced by QCD instantons in deep inelastic scattering are discussed.
First instanton searches and the prospects for their experimental
discovery at HERA are presented.}

\normalsize\baselineskip=15pt
\setcounter{footnote}{0}
\renewcommand{\thefootnote}{\alph{footnote}}

\section{Introduction}                      
Since a long time it has been recognized that
the standard model contains processes which cannot be
described by perturbation theory, and which violate
classical conservation laws like baryon number ($B$) and lepton number ($L$)
in the case of the electroweak interaction,
and chirality ($Q_5$) in the case of
the strong interaction\cite{inst:thooft}.
Such anomalous processes are induced by
so-called instantons\cite{inst:belavin}.
The name indicates that these are non-perturbative fluctuations
that are confined to ``an instant'' in space-time, with no
corresponding free particle solutions for $t\rightarrow \pm \infty$.
The interest in instantons
remained somewhat academic, as observable effects
were predicted to exist only at extremely
high energies of $\order{10^5~\TeVx}$,
until it was discovered that
their exponential suppression is much reduced
by the emission of gauge bosons\cite{inst:ringwald}.
In electroweak theory with massive gauge bosons
still rather high energies of $\order{\gtrsim 10~\TeVx}$ would be required,
but not so in QCD with massless gluons and strong coupling.
Instanton effects could play a r\^{o}le in QCD reactions
already at present day colliders.
Deep inelastic $ep$ scattering at HERA is particularly interesting,
because the virtuality of the photon probe \Qsq provides a
hard scale for the instanton subprocess,
which is needed for
theoretically sound
predictions\cite{inst:balitsky1,inst:vladimir,inst:yaroslavl}.
Instanton effects have not yet been observed in nature.
Their experimental discovery would be of fundamental
significance for particle physics.

In this report a short
introduction to the basic theoretical ideas will be given.
Instanton phenomenology in deep inelastic scattering (DIS) will
be discussed, covering cross sections and event topologies.
Finally, prospects for instanton searches and
first results from the analysis of HERA data
will be presented.

\section{Instanton theory}                  

\begin{wrapfigure}{r}{6cm}
\mbox{
   \epsfig{file=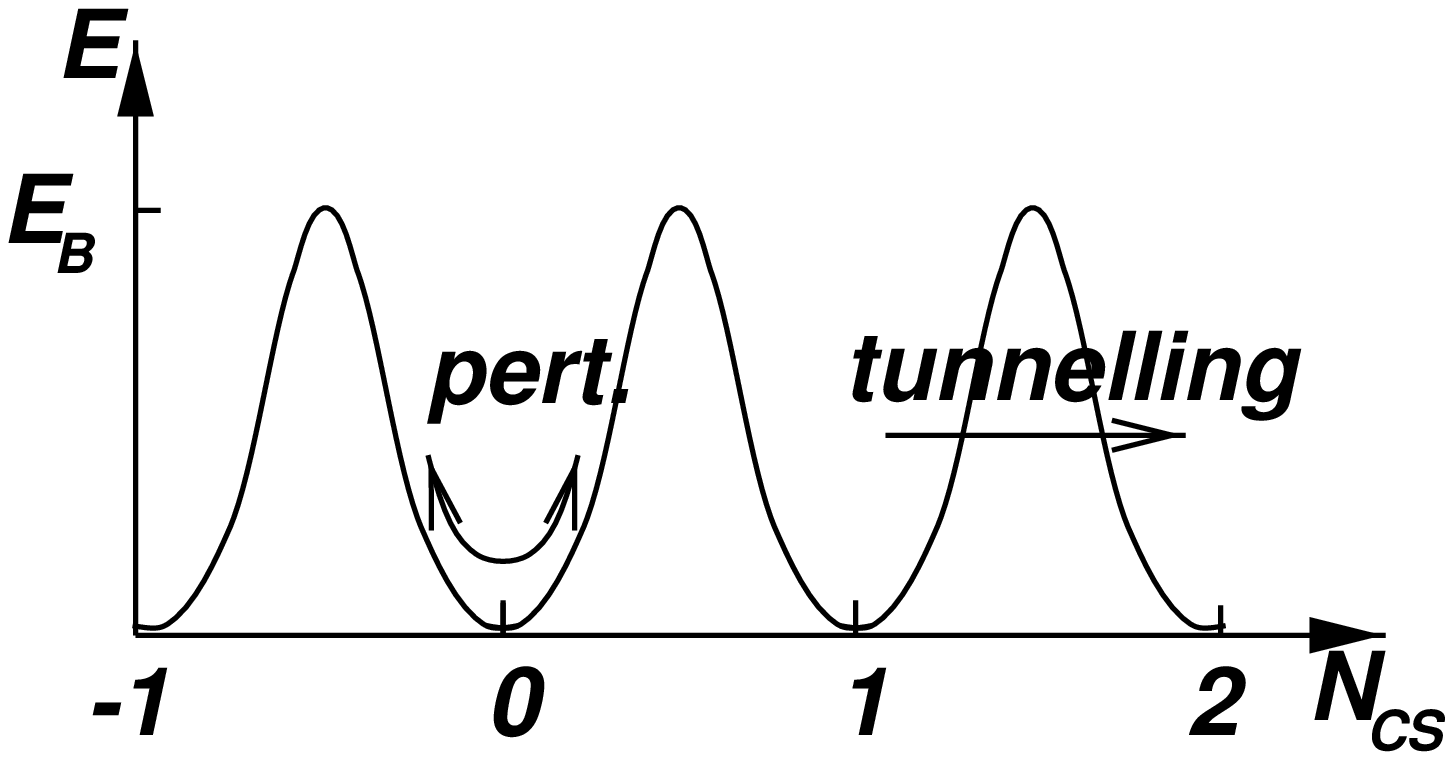,%
      width=5cm,bbllx=93pt,bblly=283pt,bburx=516pt,bbury=511,clip=}
     }
   \bild{5cm}{The structure of the vacuum.
                      Instanton solutions represent tunnelling
                      transitions between topological inequivalent
                      minima, which cannot be reached perturbatively.}
   \label{structure}
\end{wrapfigure}

Instantons originate from the non-trivial topological
structure of the vacuum in non-Abelian
gauge field theories, where
the vacuum is degenerate in
the Chern-Simons number \ncs.
\ncs is defined as an integral over the gauge
fields $A^a_\mu$ with coupling $g$,
\begin{equation}
  \ncs := \frac{g^2}{16\pi^2} \int \dd^3x \epsilon_{ijk}
         \left( A_i^a \partial_j A^a_k -
               \frac{g}{3} \epsilon_{abc} A_i^a A_j^b A_k^c \right).
\end{equation}
Neighbouring vacua have the same (minimal) potential energy,
but differ in
their topological winding number \ncs, and
are separated by a potential barrier of height
$E_B$ \mbox{(fig.~\ref{structure})}.
The usual perturbative expansion of the scattering amplitudes
in the coupling constant $\alpha$
around {\em one} minimum (fig.~\ref{structure}),
conveniently represented by a series of Feynman graphs,
does not allow for transitions between neighbouring
minima.
They may however occur classically when the energy $E$ is large enough
$E>E_B$, or by quantum mechanical tunnelling when $E<E_B$,
corresponding to so-called instanton solutions
of the classical field equations.
The transition amplitude for the
instanton--induced tunnelling process
is exponentially suppressed
$\propto \exp(-4\pi/\alpha)$, a very small number.


\begin{figure}[htb]
   \centering
\begin{picture}(0,0) \put(0,0){{\bf a)}} \end{picture}
   \epsfig{file=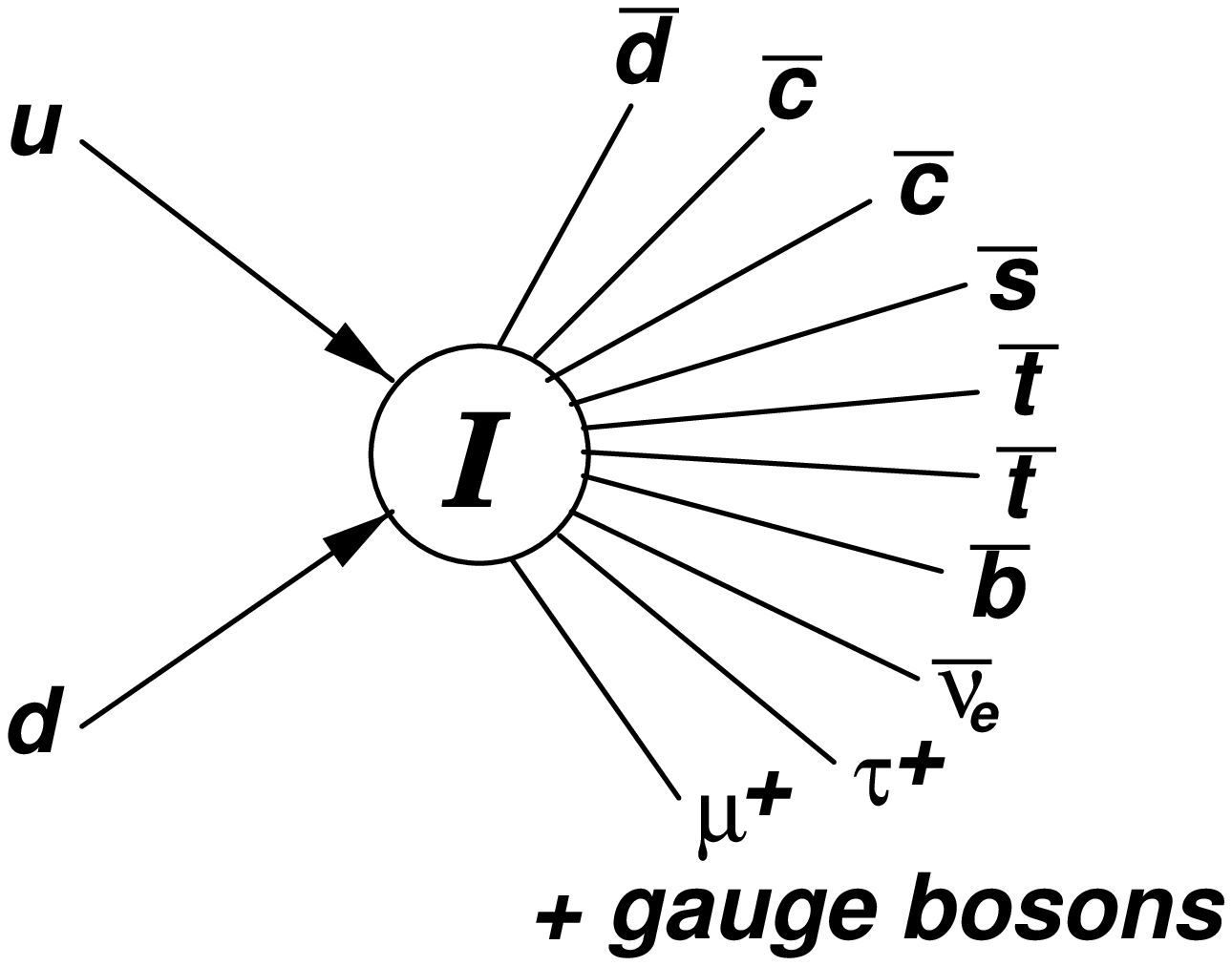,%
      width=6.0cm,bbllx=101pt,bblly=220pt,bburx=508pt,bbury=563,clip=}
   \hspace{1.5cm}
\begin{picture}(0,0) \put(0,0){{\bf b)}} \end{picture}
   \epsfig{file=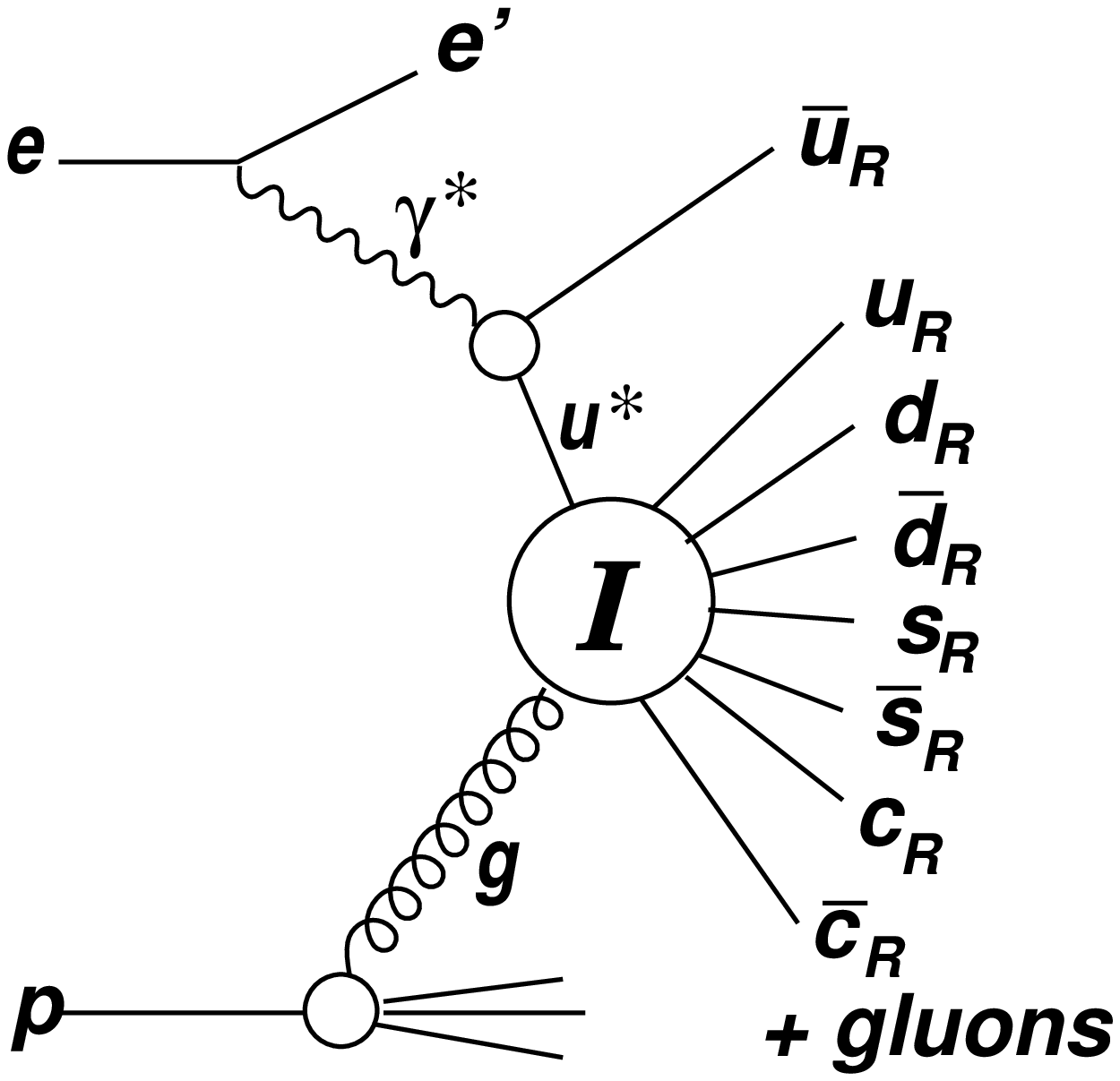,%
      width=6.0cm,bbllx=101pt,bblly=220pt,bburx=508pt,bbury=591,clip=}
\vspace{0.3cm}
   \fcaption{
           {\bf a)} the electroweak interaction with $\Delta (B+L)=-6$ and in
             {\bf b)} the strong interaction with $\Delta Q_5 = 8$.}
   \label{instviol}
\end{figure}

In the electroweak theory, the minimal barrier height is
$E_B  \approx m_W/\alpha_w = \order{10 \TeVx}$.
Instanton transitions between vacua separated by $\Delta \ncs$
(see fig.~\ref{instviol}a for an example)
would violate baryon ($B$) and lepton numbers ($L=L_e+L_\mu+L_\tau$)
according to
\begin{equation}
 \Delta (B+L) = -2~ n_{\rm gener.} \cdot\Delta \ncs,
\end{equation}
but respect
\begin{equation}
  \Delta (B-L)=0 \hspace{2cm}
  \Delta L_e = \Delta L_\mu = \Delta L_\tau = \Delta B /3.
\end{equation}
 $n_{\rm gener.}=3$ is the number of fermion generations.

In instanton induced QCD reactions
(see fig.~\ref{instviol}b)
chirality is violated.
The chirality $Q_5$ is the difference between
the number of left- and right-handed fermions, $Q_5 = \#L -\#R$.
For $n_f$ active quark flavours, the selection rule is
\begin{equation}
 \Delta Q_5 = 2~ n_f \cdot \Delta \ncs.
\end{equation}
The minimal barrier height
is given by the hard scale of the process, e.g.
$E_B=\order{Q}$ for DIS\cite{inst:vladimir}.
The exponential suppression is less
severe than in the electroweak case, because $\alpha_s \gg \alpha_w$.

\section{Instantons at HERA}                
\begin{figure}[h]
\begin{tabular}{ll}
\mbox{
 \epsfig{file=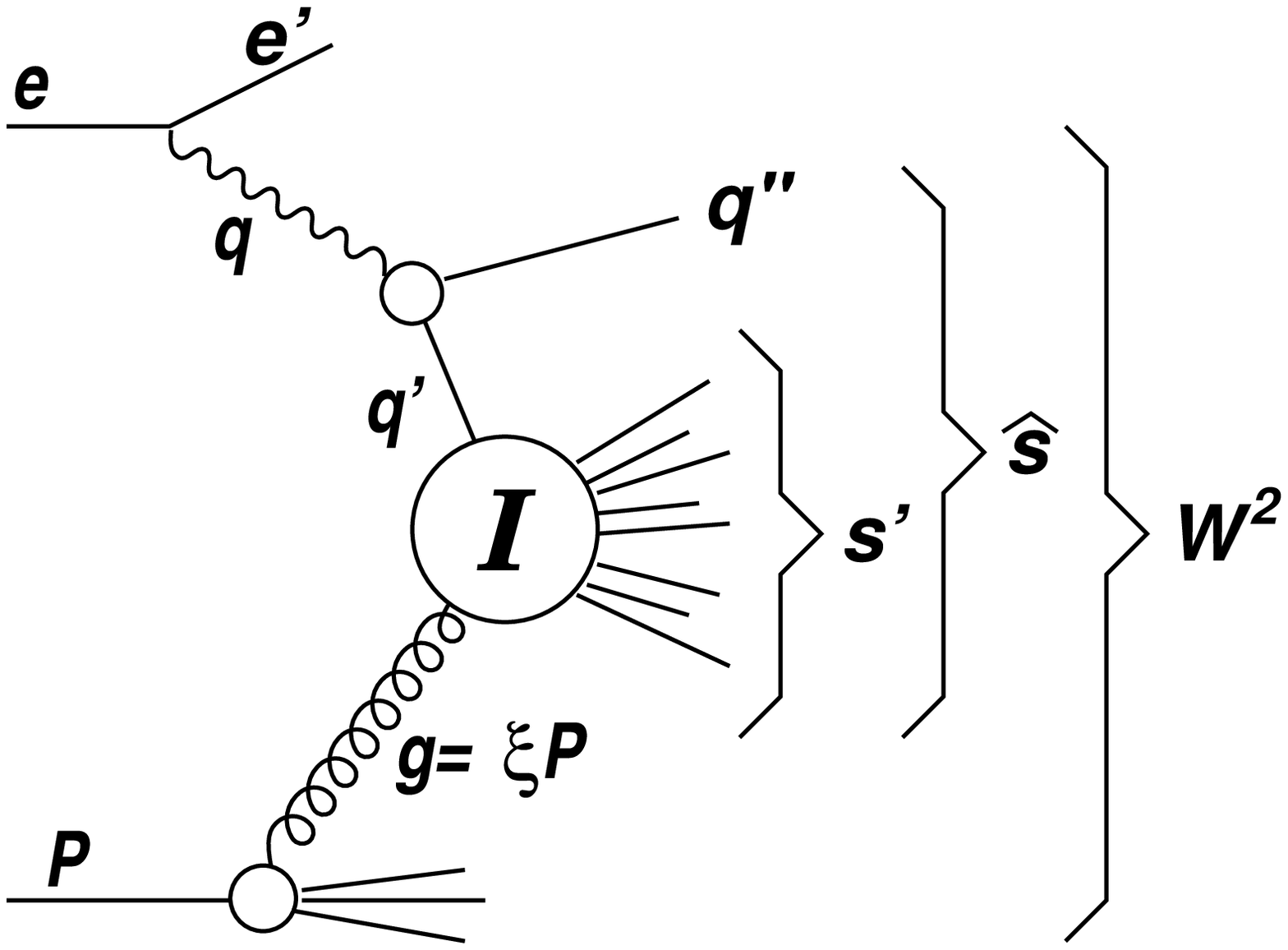,width=7.0cm,%
 bbllx=70pt,bblly=215pt,bburx=534pt,bbury=569,clip=}
}
&
\begin{tabular}{l}
 \vspace{-5.5cm} \\
 DIS variables: \\
   $\Qsq := - q^2 $ \\
   $ x := \Qsq / (2 P \cdot q) $ \\
   $ W^2 :=(q+P)^2 = \Qsq (1 - x)/x$ \\
   $ \hat{s} := (q+g)^2$ \\
   $ \xi = x (1+\hat{s}/Q^2)$ \\ \\
 Variables of instanton subprocess: \\
 $\qprimesq:=- q'^2 $ \\
 $x':= \qprimesq /(2 \; g \cdot q' ) $ \\
 $s':= (q'+g)^2 = \qprimesq ( 1 - \xprime )/ \xprime $
\end{tabular}
\end{tabular}
\vspace{0.3cm}
   \fcaption{Kinematics of instanton induced processes in DIS.
    The labels denote the 4-vectors of the particles.
    A virtual photon $\gamma^\ast$ (4-momentum $q=e-e'$)
    emitted from the incoming electron fuses with
    a gluon (4-momentum $g$) from the proton (4-momentum $P$).
    The gluon carries a fraction $\xi$ of the proton momentum.
    The virtual quark $q^*$ entering
    the instanton subprocess has 4-momentum $q^\prime$, and the outgoing
    quark from the $\gamma^\ast\rightarrow q\ol{q}$ splitting
    has 4-momentum $q^{\prime\prime}$. The invariant masses squared of the
    $\gamma^\ast g$ and $q^*g$ systems are \shat~ and $s'$.
    $W$ is the
    invariant mass
    of the total hadronic system (the $\gamma^\ast p$ system).
    $0 \leq x \leq x/\xi \leq x' \leq 1$ holds.
    For completeness, we
    note $y:= (Pq)/(Pe) = Q^2/(sx)$, where $s=(e+P)^2$ is the $ep$
    invariant mass squared.
    }
   \label{fig:inst}
\end{figure}

In recent years, it has been
realized\cite{inst:vladimir,inst:yaroslavl,inst:moch,inst:schremppdis96,%
inst:ringwalddis97,inst:schremppdis97}
that quantitative
calculations are possible for processes induced
by QCD instantons in DIS due
to the presence of a hard
scale, $Q^2$.
In DIS, events due to QCD instantons $I$ (and anti-instantons \ol{I})
are predominantly produced
in a photon-gluon fusion
processes\footnote{Quark initiated processes have not yet been considered.
Due to the large gluon content of the proton
in the HERA domain at small $x$, they
are expected to be of minor importance.
In addition, they are expected to be suppressed by $\order{\alpha_s^2}$
with respect to the gluon initiated processes.}
~(Fig.~\ref{fig:inst})
\begin{equation}
  \gamma^\ast+g \stackrel{I}{\rightarrow}
            \sum_{n_f} (\ol{q}_R+q_R)+n_g g
 \hspace{1cm}
  \gamma^\ast+g \stackrel{\overline{I}}{\rightarrow}
            \sum_{n_f} (\ol{q}_L+q_L)+n_g g.
\end{equation}
In each event, quarks and antiquarks of all $n_f$ active flavours are found,
with $n_g$ gluons in addition.

The kinematics is depicted in fig.~\ref{fig:inst}.
The DIS variables Bjorken $x$ and $Q^2$
can be measured from the scattered electron, $q=e-e'$.
A measurement of the other variables is more challenging.
A measurement of the invariant
mass of the hadronic system, excluding the remnant, would determine \shat.
If the outgoing ``current jet'' could be identified and measured,
it's 4-momentum $q^{\prime\prime}$ would determine
$q'=q-q^{\prime\prime}$, and thus the
variables $x'$ and $Q'^2$ which characterise the instanton subprocess.
In practice, when not all of the five independent invariants
(for example ${x,Q^2,x',Q'^2,\hat{s}}$) can be measured, they are being
integrated over.

The instanton induced cross section is given by a convolution of the
probability to find a gluon in the proton $P_{g/p}$,
the probability that the virtual photon splits
into a quark-antiquark pair in the instanton background
$P^{(I)}_{q^*/\gamma^*}$, and
the cross section
$\sigma^{(I)}_{q^*g}(\xprimex,\qprimesqx)$ of the instanton
subprocess\cite{inst:yaroslavl,inst:moch}.
Multi-gluon emission enhances the cross section\cite{inst:ringwald}
\begin{equation}
\sigma_{q^*g;n_g}^{(I)} \propto \frac{1}{n_g!}
                 \left(\frac{1}{\alpha_s}\right)^{n_g}
                \exp(-4\pi/\alpha_s).
\end{equation}
The cross section of the instanton induced subprocess is
then\cite{inst:yaroslavl}:
\begin{equation}
 \sigma_{q^*g}^{(I)}(\xprimex,\qprimesqx)
 = \sum_{n_g=0}^{\infty}\sigma_{q^*g;n_g}^{(I)}
 \approx
 \frac{\Sigma(\xprimex)}{\qprimesqx}
{ \left(\frac{4 \pi}{\alpha_s(\mu(\qprimex))}\right)}^{\frac{21}{2}}
 {\rm exp} \left(\frac{- 4 \pi}{\alpha_s(\mu(\qprimex))} F(\xprimex) \right).
\label{eq:cross}
\end{equation}
\begin{wrapfigure}{r}{8cm}
\vspace{-0.8cm}
\mbox{
   \epsfig{file=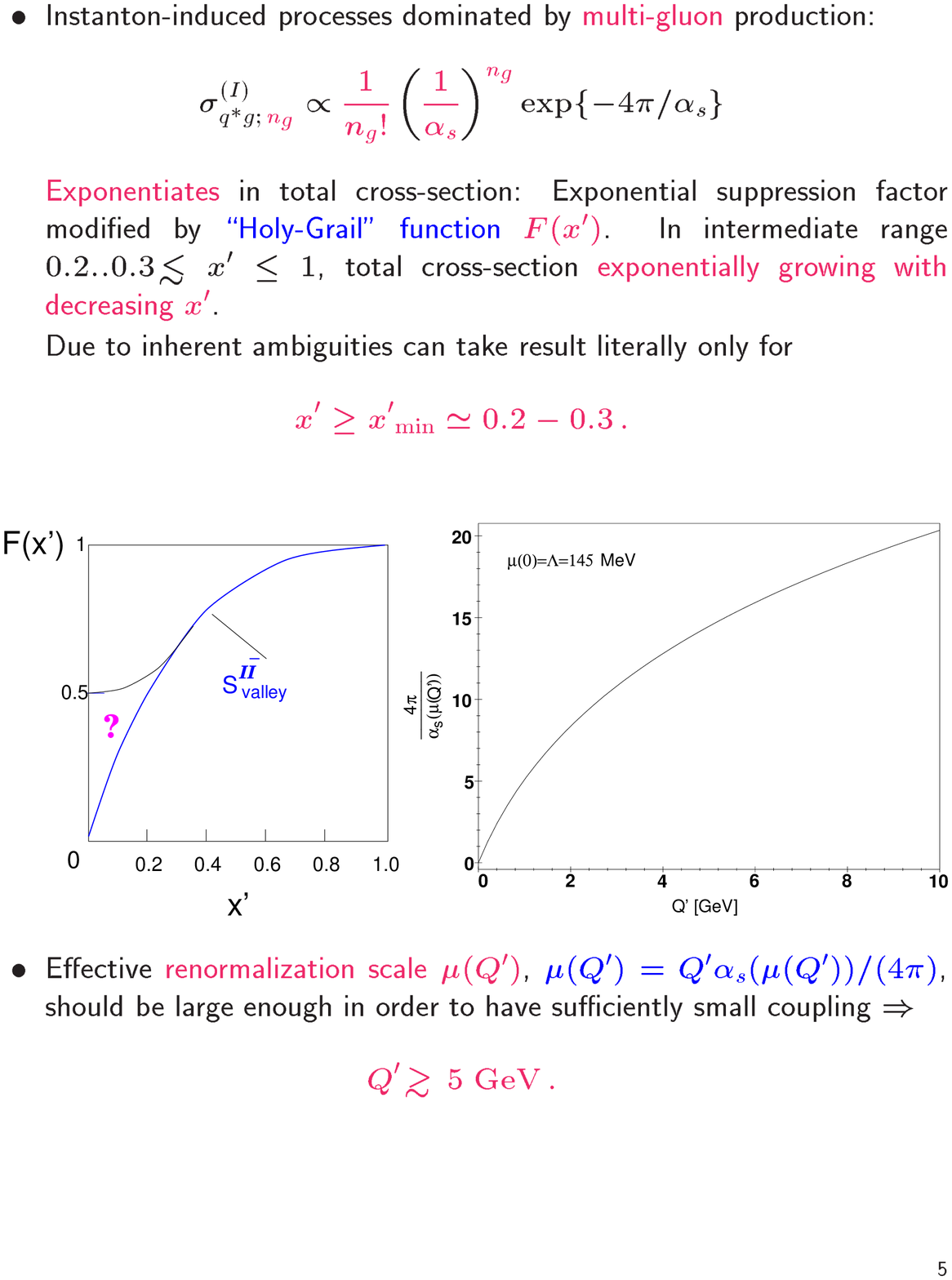,%
      width=5cm,bbllx=12pt,bblly=253pt,bburx=250pt,bbury=484,clip=}
     }
   \bild{8cm}{The holy grail function $F(x')$ \cite{inst:yaroslavl}.
              For small $s'$ ($x'\approx 1$), instanton perturbation
              theory is applied. The calculation with the valley
              method matches smoothly with the perturbative result.}
   \label{grail}
\end{wrapfigure}
It depends critically on the functions
$F(\xprimex)$ (called the ``holy grail'' function),
which modifies the exponent in the suppression factor
$\exp(-4 \pi /\alpha_s)$,
and on $\Sigma(\xprimex)$, which depends on $F(\xprimex)$.
There exists also a scale dependence due to the choice
of the renormalization scale $\mu(\qprimex)$.

$F(\xprimex)$
can be estimated reasonably well  (see fig.~\ref{grail})
for \xprime not
too small,
$\xprime \gtrsim 0.2$ \cite{inst:yaroslavl}.
The extrapolation to lower
values of \xprime is unreliable due to
inherent ambiguities.
In addition, multi-instanton effects
should be avoided
by limiting the instanton size $\rho_I$
(the spatial region occupied during the interaction)
to $\rho_I<2 {\rm ~GeV}^{-1}$ with
a cut-off
$\qprimesq \gtrsim 25 \GeVsq$ \cite{inst:yaroslavl,inst:moch}.
That requirement ensures also that $\alpha_s(\mu(\qprimex))$
stays small enough to apply instanton perturbation theory.

\begin{figure}[htb]
   \centering
 \begin{picture}(0,0) \put(0,0){{\bf a)}} \end{picture}
   \epsfig{file=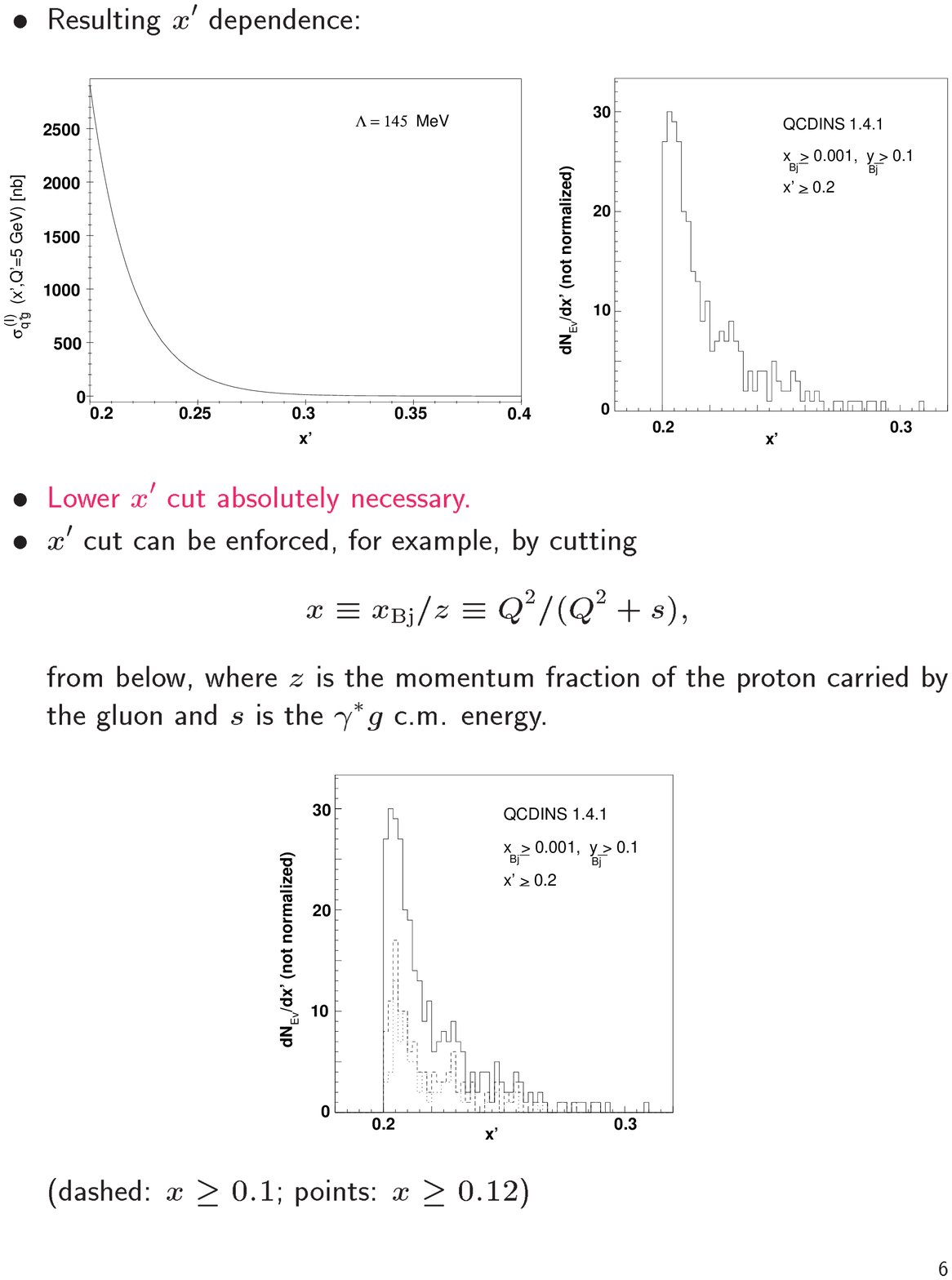,%
      width=7.4cm,bbllx=16pt,bblly=520pt,bburx=330pt,bbury=747,clip=}
 \begin{picture}(0,0) \put(0,0){{\bf b)}} \end{picture}
   \epsfig{file=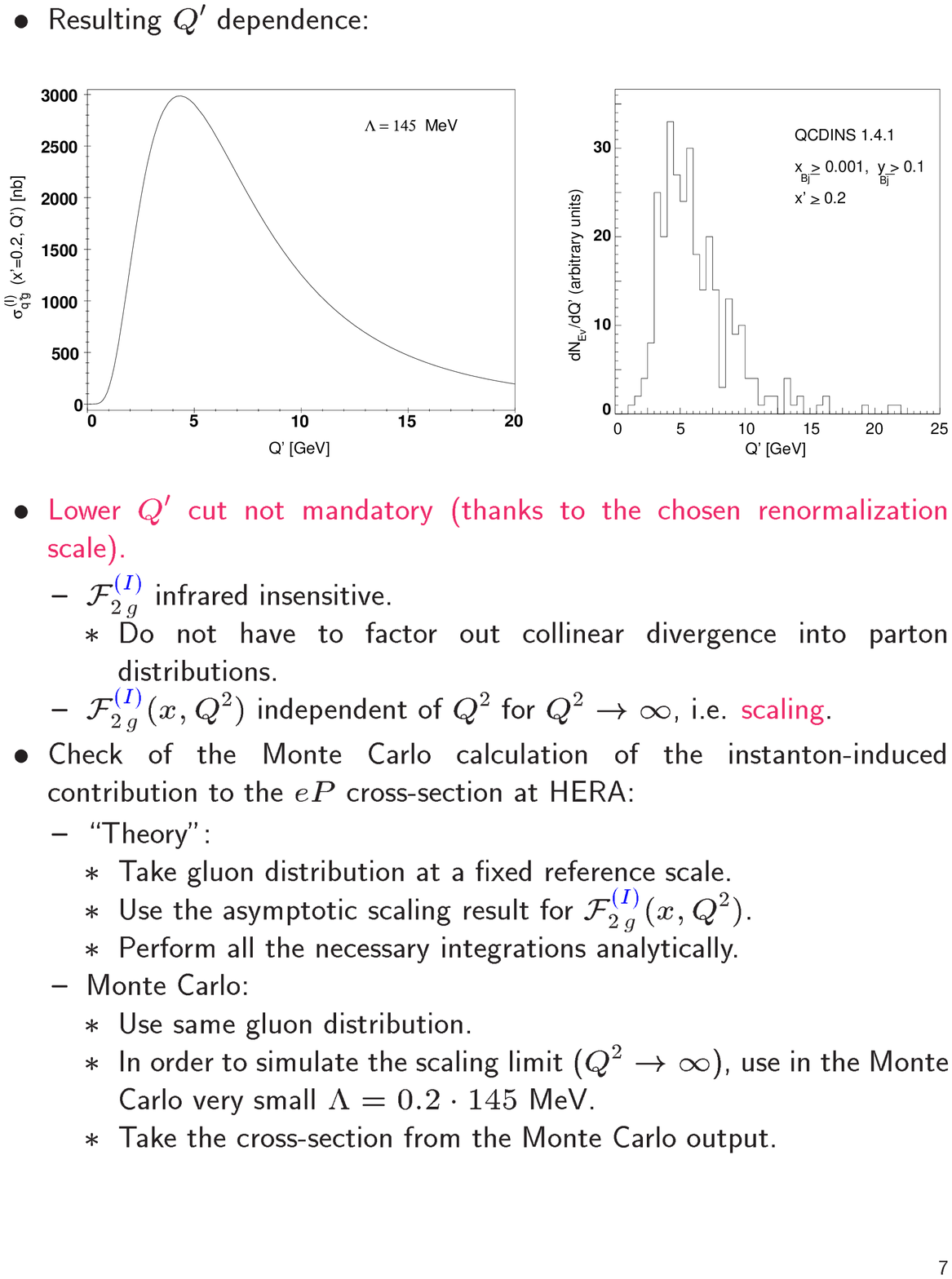,%
      width=7.4cm,bbllx=16pt,bblly=520pt,bburx=330pt,bbury=747,clip=}
   \fcaption{
                The instanton subprocess
                cross section\cite{inst:ringwalddis97}
                for $q^\ast g \rightarrow$~hadrons
                as a function of
              {\bf a)} $x'$ for $Q'=5 \GeV$ and
              {\bf b)} $Q'$ for $x'=0.2$.}
   \label{sigsub}
\end{figure}

\begin{figure}[htb]
   \centering
   \epsfig{file=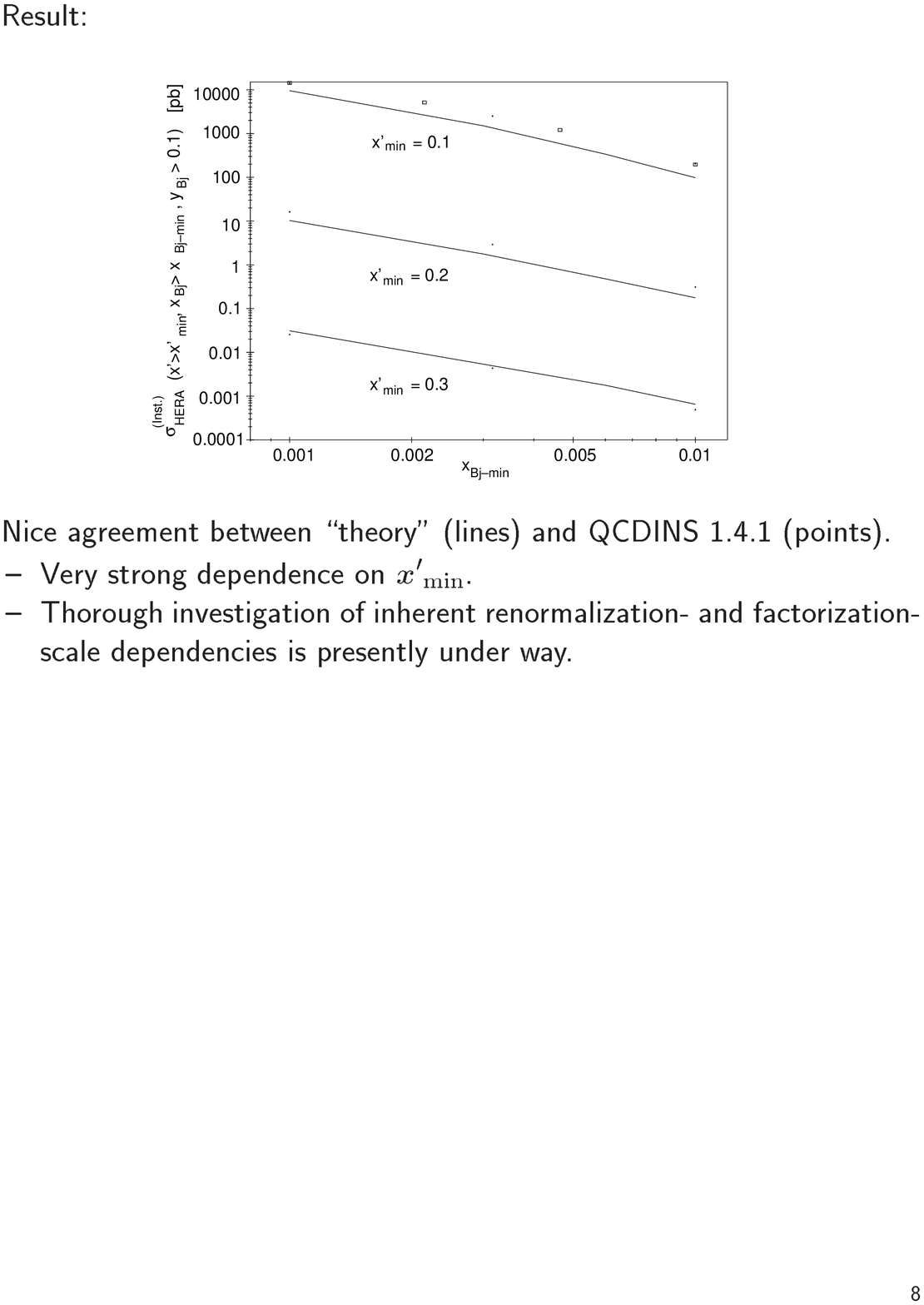,%
      width=10cm,bbllx=126pt,bblly=516pt,bburx=465pt,bbury=750,clip=}
\vspace{0.3cm}
   \fcaption{
                The instanton induced DIS cross
                section\cite{inst:ringwalddis97}
                $ep \rightarrow e'X$ integrated
                over Bjorken $x>x_{\rm min}$, $y>0.1$, and over
                the regions $Q'>5\GeV$ and $x'>x'_{\rm min}$
                as indicated.}
   \label{sigi}
\end{figure}

The resulting instanton induced subprocess cross section
$\sigma_{q^*g}^{(I)}(\xprimex,\qprimesqx)$
(see fig.~\ref{sigsub})
is peaked at $Q'\approx 5 \GeV$
and exponentially grows with decreasing \xprimex.
The integrated instanton induced
$ep$ DIS cross section (see fig.~\ref{sigi})
is sizeable;
for $x>0.001$ and $x'>0.2$ it is of \order{10~\pbx}.
The cross section is approximately
scaling (depends only on $x$, not on \Qsq for large \Qsq) \cite{inst:moch}.
It grows towards small $x$, and increases dramatically
when the lower $x'$ cut-off is relaxed.
Eventually
higher order instanton effects have to dampen the growth of
the cross section.

Two kinematic regions have to be distinguished.
For $x'>0.2$ the predictions are relatively safe,
allowing the instanton theory to be tested. Either instantons are
discovered at the predicted level
-- including the substantial theoretical uncertainties, which still need
   to be quantified --,
or the theory has to be revised.
For $x'<0.2$ the cross section presumably continues to grow, but the
extrapolation is extremely uncertain.
For a discovery, this is the favourable region
due to the large cross section.
A negative result however cannot be turned against the theory,
it would rather restrict the unknown behaviour of $F(x')$ at small $x'$.
Most promising is the kinematic region of small Bjorken-$x$,
because both the total DIS cross section and the predicted
fraction of instanton induced events increase towards small $x$
(see fig.~\ref{dpdn}b).

\section{Experimental signatures}            
In the theoretically safe region, $x'>0.2$,
the expected fraction of instanton events in all DIS events is
of \order{10^{-3} - 10^{-4}} (compare fig.\ref{dpdn}b),
too small to be detected in
inclusive cross section measurements
(i.e. the structure function $F_2$).
Instead, dedicated searches for the characteristic features
of instanton events in the hadronic final state have to be performed.
A Monte Carlo generator (QCDINS\cite{inst:qcdins})
to simulate the hadronic final state of
instanton events in DIS is available.
In general, the event shape predictions are more stable than the
rate predictions, because poorly known factors cancel.
The instanton event properties can be contrasted with predictions
from event generators for normal DIS events
(ARIADNE\cite{mc:ariadne}, LEPTO\cite{mc:lepto} and HERWIG\cite{mc:herwig})
which give an overall satisfactory
description of the DIS final state properties\cite{mc:heratune}.

In the $q^\ast g$
rest frame
$2 n_f-1$ quark and antiquarks
and $n_g$ gluons are emitted isotropically
from the instanton subprocess.
$n_g$ is Poisson distributed with\cite{inst:moch,inst:ringwalddis97}
\begin{equation}
  \av{n_g} \approx \frac{2\pi}{\alpha_s} x' (1-x')
  \frac{\dd F(x')}{\dd x'}.
\end{equation}
After hadronization, this leads to a spherical system with a high
multiplicity of hadrons, depending mainly
on the available centre of mass energy
$\sqrt{s'} = Q'\sqrt{1/x'-1}$.
For a typical situation ($x'=0.2,Q'=5\GeV \Rightarrow \sqrt{s'}=10~\GeVx$),
$\av{n_g} = \order{2}$.
About $n_p=10$ partons
and $n=20$ hadrons are expected. The expected parton momentum spectrum
is semi-hard\cite{inst:vladimir} with transverse momentum
$\av{p_T} \approx (\pi/4)(\sqrt{s'}/\av{n_p})$.

Hadronic final state properties are conveniently
being studied in the centre of mass system (CMS)
of the incoming proton and the virtual boson, i.e. the CMS of
the hadronic final state.
Longitudinal and transverse quantities are calculated
with respect to the virtual boson direction (defining the $+z$ direction).
The pseudorapidity
$\eta$ is defined as $\eta= - \ln \tan (\theta/2) $, where
$\theta$ is the angle with respect to the virtual
photon direction.
When boosted to the CMS,
the hadrons emerging from the instanton subprocess occupy a
band in pseudorapidity of half width $\Delta \eta \approx 1$,
which is homogeneously populated in azimuth \cite{inst:vladimir}.

The characteristics of instanton events by which they
can be distinguished from normal DIS events are therefore:
high multiplicity with large transverse energy; spherical
event configuration (apart from the current jet);
and the presence of all flavours (twice!)
that are kinematically allowed in each event.
One would therefore look for events which in addition to
the other characteristics are rich in $K^0$s, charm decays,
secondary vertices,
muons etc..
In general, the strength of instanton signals in the hadronic
final state increases somewhat towards low $x'$ and large $Q'^2$
due to the increasing ``instanton mass'' $\sqrt{s'} = Q'\sqrt{1/x'-1}$.

\begin{figure}[h]
   \centering
 \begin{picture}(0,0) \put(0,0){{\bf a)}} \end{picture}
 \epsfig{%
    file=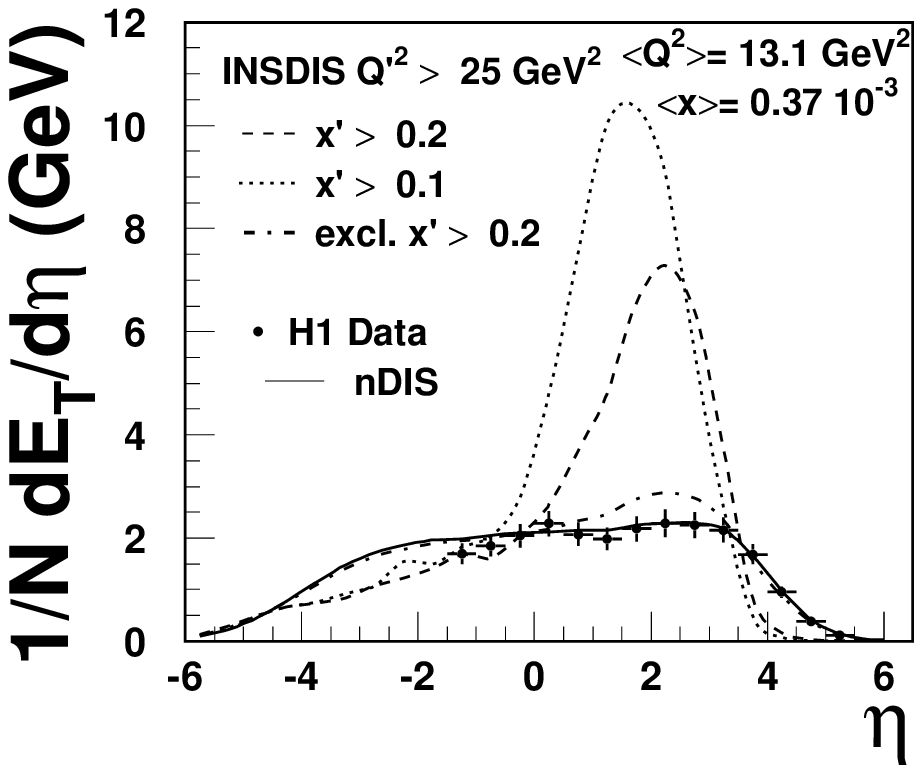,%
    width=7cm}%
 \begin{picture}(0,0) \put(0,0){{\bf b)}} \end{picture}
 \epsfig{file=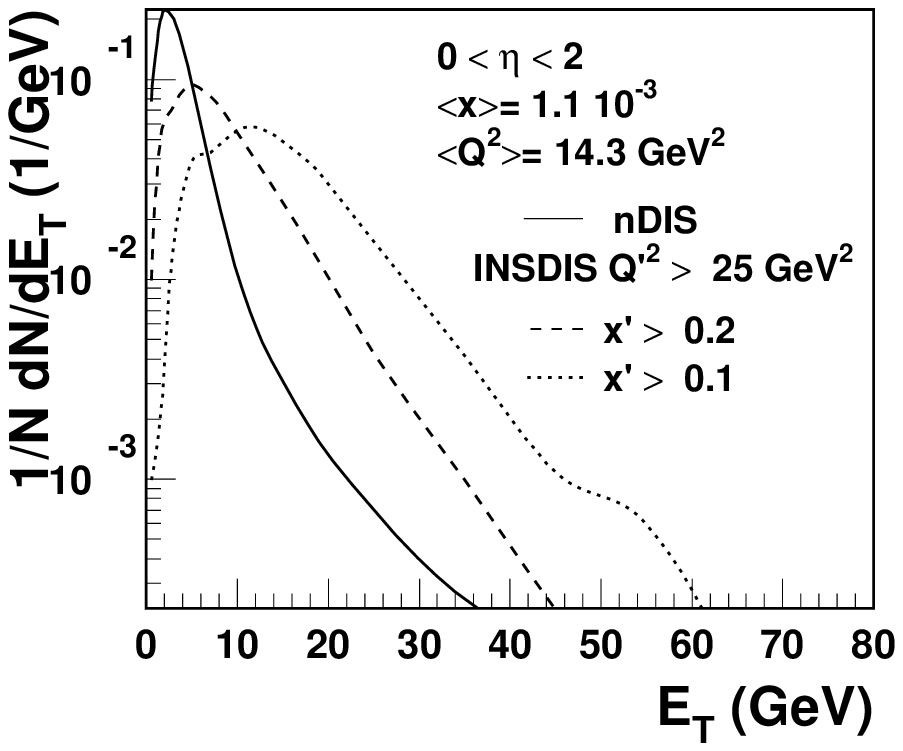,%
  width=7cm}%
\vspace{0.3cm}
   \fcaption{Transverse energy \et in the hadronic CMS \cite{inst:bounds}.
             {\bf a)} The \et flow vs. $\eta$.
             The proton remnant direction is to the left.
             The standard QCD model (nDIS=ARIADNE) and
             different instanton scenarios are confronted with
             the H1 data \cite{h1:flow3}.
             The excluded scenario \cite{inst:bounds}
             with an instanton fraction $f_I> 11.8\%$ for
             $\xprime > 0.2$ is indicated.
             {\bf b)}
             The \et distribution, where the transverse energy
             is measured in the CMS rapidity bin $0<\eta<2$,
             for two instanton scenarios, and the standard QCD model
             (nDIS=ARIADNE).
             The plots are normalized to the
             total number of events $N$ that enter the distributions.
            }
   \label{flows}
\end{figure}

The ``instanton band'' shows up in
the flow of hadronic transverse energy \et as a function of
$\eta$
(fig.~\ref{flows}a).
It's height and position depends on \xprime and \qprimesq
(and also on \xb and \Qsq).
In normal DIS events on average an \et of $2\GeV$ per $\eta$ unit
is observed. In instanton induced events, that number may go up
to $10$~\GeV for low \xprimex.
A possible search strategy could involve
the \et distribution in a selected rapidity band (fig.~\ref{flows}b),
looking for high \et events in the
tail of the distribution\cite{inst:heraws96,inst:bounds}.

\begin{figure}[htb]
   \centering
   \epsfig{file=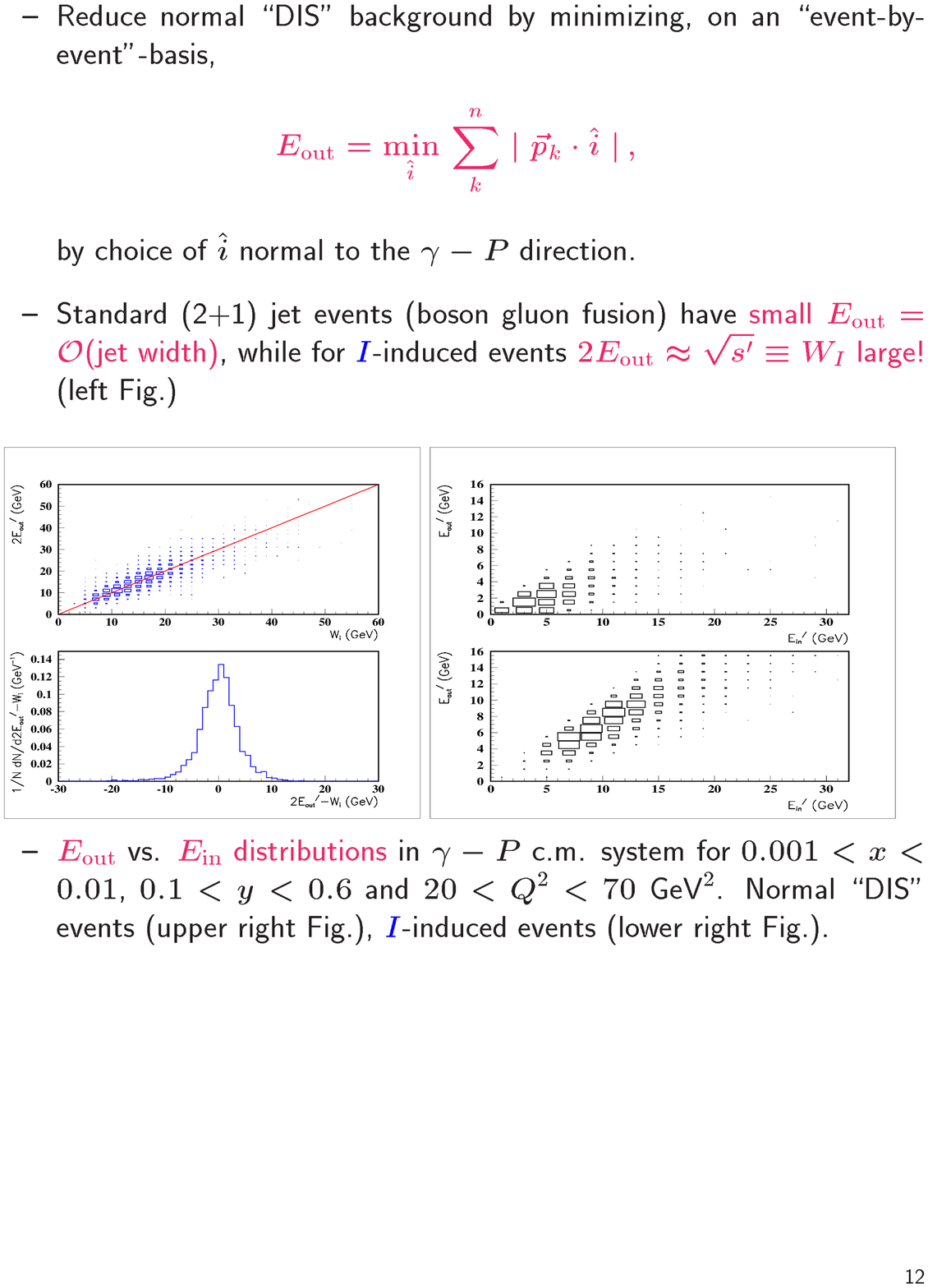,%
      width=14.5cm,bbllx=35pt,bblly=316pt,bburx=550pt,bbury=534,clip=}
   \fcaption{
             {\bf a)}
                The correlation between $2\cdot\eoutp$ and
                the ``instanton mass'', $W_I=\sqrt{s'}$ (top),
                and the resolution for $\sqrt{s'}$ that can be achieved
                (bottom)\cite{inst:heraws96}.
                The primes indicate
                additional cuts in $\eta$ to minimize higher order
                QCD radiation which may wash out the relation
                between \eout and $\sqrt{s'}$.
             {\bf b)}~$\eoutp$ vs. $\einp$ for
                normal (top, HERWIG) and instanton
                induced events (bottom, QCDINS)\cite{inst:heraws96}.
                Both distributions are taken in the hadronic CMS
                for events with $0.001<x<0.01$, $0.1<y<0.6$ and
                $20 \GeVsq < Q^2 < 70 \GeVsq$.
             }
   \label{eti}
\end{figure}

Further discrimination can be obtained\cite{inst:heraws96}
from the fact that
for instanton events the \et should be distributed isotropically,
while normal DIS events are jet-like, in particular for large \et.
One defines
\begin{equation}
\eout  := \min \sum_{i} \vec{p}_i \cdot \vec{n} \hspace{2cm}
\ein   := \sum_{i} \vec{p}_i \cdot \vec{n}'.
\end{equation}
The sum runs over all final state hadrons $i$ with momentum $\vec{p}_i$.
$\vec{n}$ is the unit vector perpendicular to the virtual photon axis which
minimizes \eout and thus defines the event plane.
$\vec{n}'$ lies in the event plane and
is normal to both $\vec{n}$ and the virtual photon axis.
It is easy to show that for an ideal isotropic ``instanton decay'',
$\eout=\sqrt{s'}/2$ \cite{inst:heraws96}.
The ``instanton mass'' $\sqrt{s'}$ can thus be
reconstructed experimentally (fig.~\ref{eti}a).
Normal DIS events, either ``1+1'' or ``2+1'' jet events (the +1 refers
to the unobserved proton remnant) are contained in the event plane,
$\eout\ll\ein$, in contrast to instanton events with $\eout \approx \ein$
(see fig.~\ref{eti}b).

Instanton events are characterized by
a large particle density localized in rapidity. In normal
DIS events there are about 2 charged particles per unit of
pseudorapidity\cite{h1:pt}, rather uniformly distributed in $\eta$.
For a low \xprime cut-off, that number goes up to 10
in the peak of the instanton band\cite{inst:bounds}.
Very sensitive to instanton events is the charged particle
multiplicity distribution\cite{inst:bounds}, see fig.~{\ref{dpdn}a.
A significant fraction of the instanton events would
lead to charged multiplicities which are very
unlikely to be found in normal DIS events.
Furthermore, particle-particle correlation functions should
be influenced by instanton effects\cite{inst:kuvshinov}.

\begin{figure}[htb]
   \centering
\begin{picture}(0,0) \put(0,0){{\bf a)}} \end{picture}
\epsfig{
    file=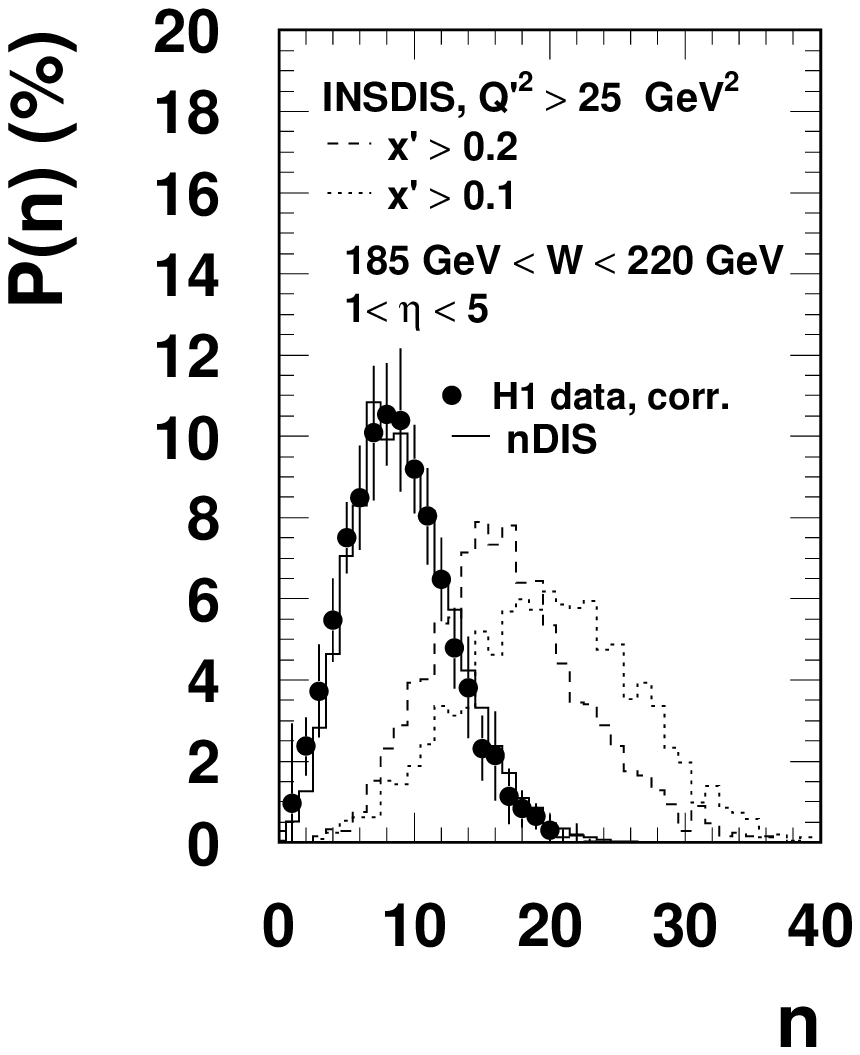,
    width=6cm,bbllx=62pt,bblly=436pt,bburx=310pt,bbury=742,clip=}
\begin{picture}(0,0) \put(0,0){{\bf b)}} \end{picture}
\epsfig{width=7.5cm,
   file=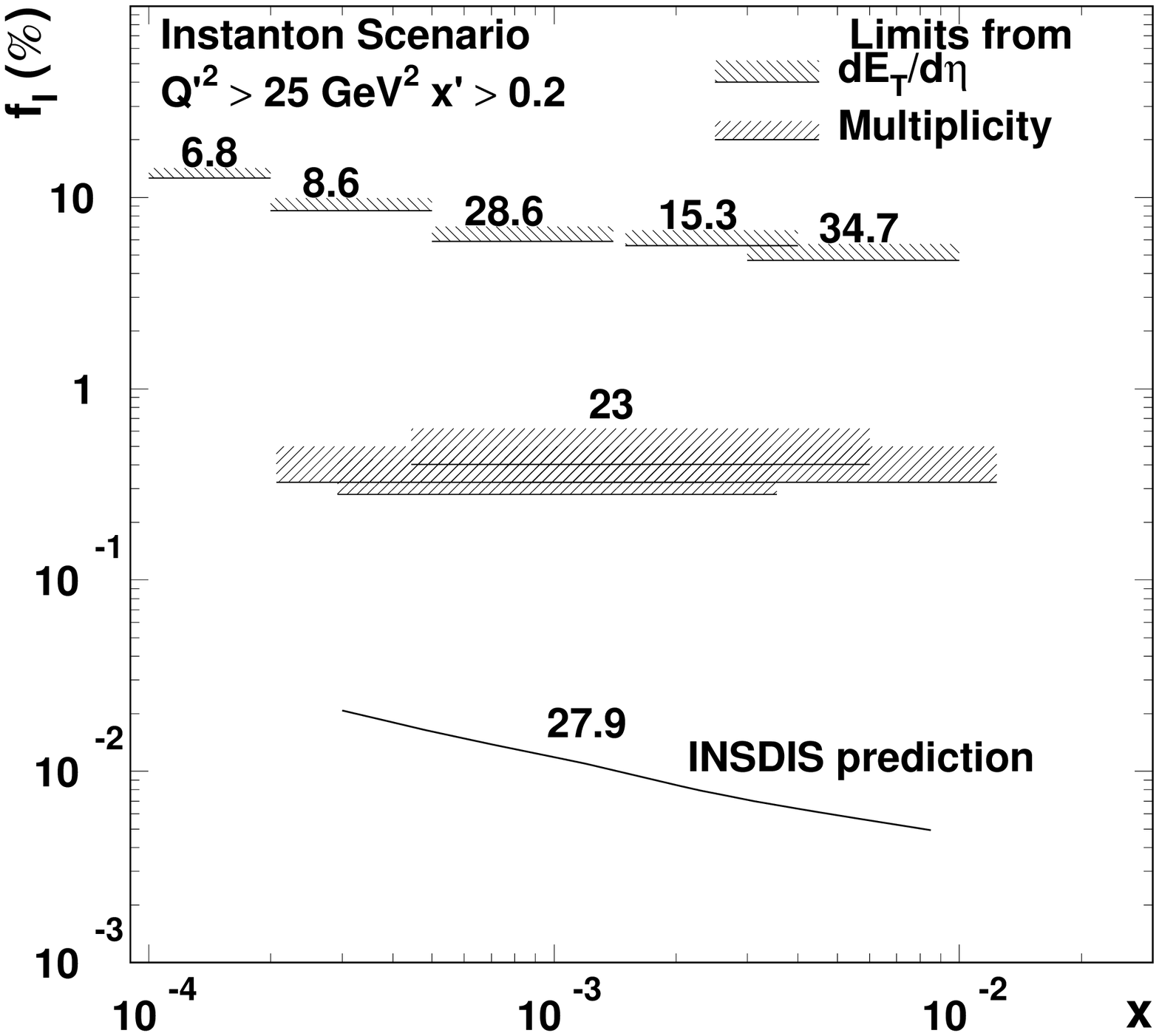,
   bbllx=0pt,bblly=-60pt,bburx=554pt,bbury=518,clip=}
\vspace{0.3cm}
   \fcaption{
            {\bf a)}  The
              probability distribution $P(n)$ of the charged particle
              multiplicity $n$
              from the CMS pseudorapidity range $1<\eta<5$ for
              events with $185 \GeV < W < 220~\GeVx$.
              Shown are the unfolded H1 data \cite{h1:mult},
              the expectation from a standard
              DIS model (nDIS=ARIADNE)),
              and the predictions for instanton events with
              different cut-off scenarios \cite{inst:bounds}.
      {\bf b)}
      The maximally allowed fraction $f_{\rm lim}$
      of instanton induced events in DIS for
      $\qprimesq > 25$\GeVsq and $\xprime > 0.2$
      from transverse energy flows and the multiplicity distribution
      as function of $x$ \cite{inst:bounds}.
      Regions above the lines are excluded
      at 95\% C.L..
      The numbers give the average \Qsq values in \GeVsq for
      the $x$ bins.
      The theory prediction, calculated with QCDINS \cite{inst:qcdins},
      for $10 \GeVsq < \Qsq < 80$~\GeVsq
      is superimposed (full line, label INSDIS).
            }
   \label{dpdn}
\end{figure}

\section{Searches for instanton processes}  
The fact that instanton events look very different from the expectation for
standard QCD events can be exploited to search for instanton signals
in the HERA data. One strategy is to compare the shape of
hadronic final state distributions to the expectation from
standard QCD events (nDIS) with an admixture of instanton events (INSDIS)
of fraction \finst. In case the measured distribution agrees with
the standard QCD expectation, a limit on the fraction of instanton
induced events in DIS $\finst<\flim$ can be set.
The caveat of this method is that one has to make an assumption
on what standard QCD looks like.
In particular at small $x$ that issue is
under debate
\cite{lowx:ringcarli,lowx:ringgrindh,mk:madrid}.
There exists a danger that an instanton effect is tuned or explained
away by stretching the standard QCD predictions
by generator tuning, introducing for example BFKL effects etc..
A good understanding of standard QCD will be crucial for the
positive identification of instanton effects.

In the first search for instanton events\cite{h1:k0} an anomalous
$K^0$ yield has been looked for.
For $x>10^{-3}$ about 0.12 $K^0$
mesons (including $\ol{K^0}$) have been measured per event and
unit pseudorapidity, with a relatively flat $\eta$ distribution.
For instanton events with $x'>0.2$,
a peaked distribution with
about 0.55 $K^0$ per event and unit $\eta$ is expected.
From the comparison with
standard QCD event generators\cite{mc:ariadne,mc:lepto,mc:herwig}
and the instanton generator\cite{inst:qcdins}
an upper limit of $f_I<\flim =6\%$ at 95\% C.L.
is obtained\cite{h1:k0}.

The charged particle multiplicity distribution $P(n)$ in high energy
reactions can often by described by a negative binominal distribution (NBD).
Also the DIS data are relatively well described by NBDs \cite{h1:mult}.
The multiplicity distribution
from the CMS interval $1<\eta<5$ for events with $W=80-115\GeV$
(corresponding to $x>0.0007$) can be parametrized
with an NBD of mean $\av{n}= 6.90 \pm 0.33$. Possible
deviations from an NBD allow for an
instanton fraction of at most \finst=2.7\% at 95\% C.L. \cite{h1:mult}.


Other measured event shapes have been systematically
analysed in terms of their sensitivity
to instanton events\cite{inst:bounds}, and their dependence on the
kinematic variables $x,Q^2,x',Q'^2$.
The most sensitive distributions were
the transverse energy flows\cite{h1:flow3},
the pseudorapidity distribution of charged particles and their
\pt spectra\cite{h1:pt}.
For example, the \et flow
has been measured over a wide range of $x$ and $Q^2$, allowing to extend
the search region down to $x=0.0001$.
From a shape analysis\cite{inst:bounds}
(see fig.~\ref{flows}),
instanton fractions \finst between 5 and 13 \%
can be excluded for $x'>0.2$ (see figs.~\ref{dpdn}b, \ref{limplane}).
For lower $x'$ the signal is more prominent, and somewhat
better limits are obtained.

\begin{figure}[tp]
 \centering
 \epsfig{%
   file=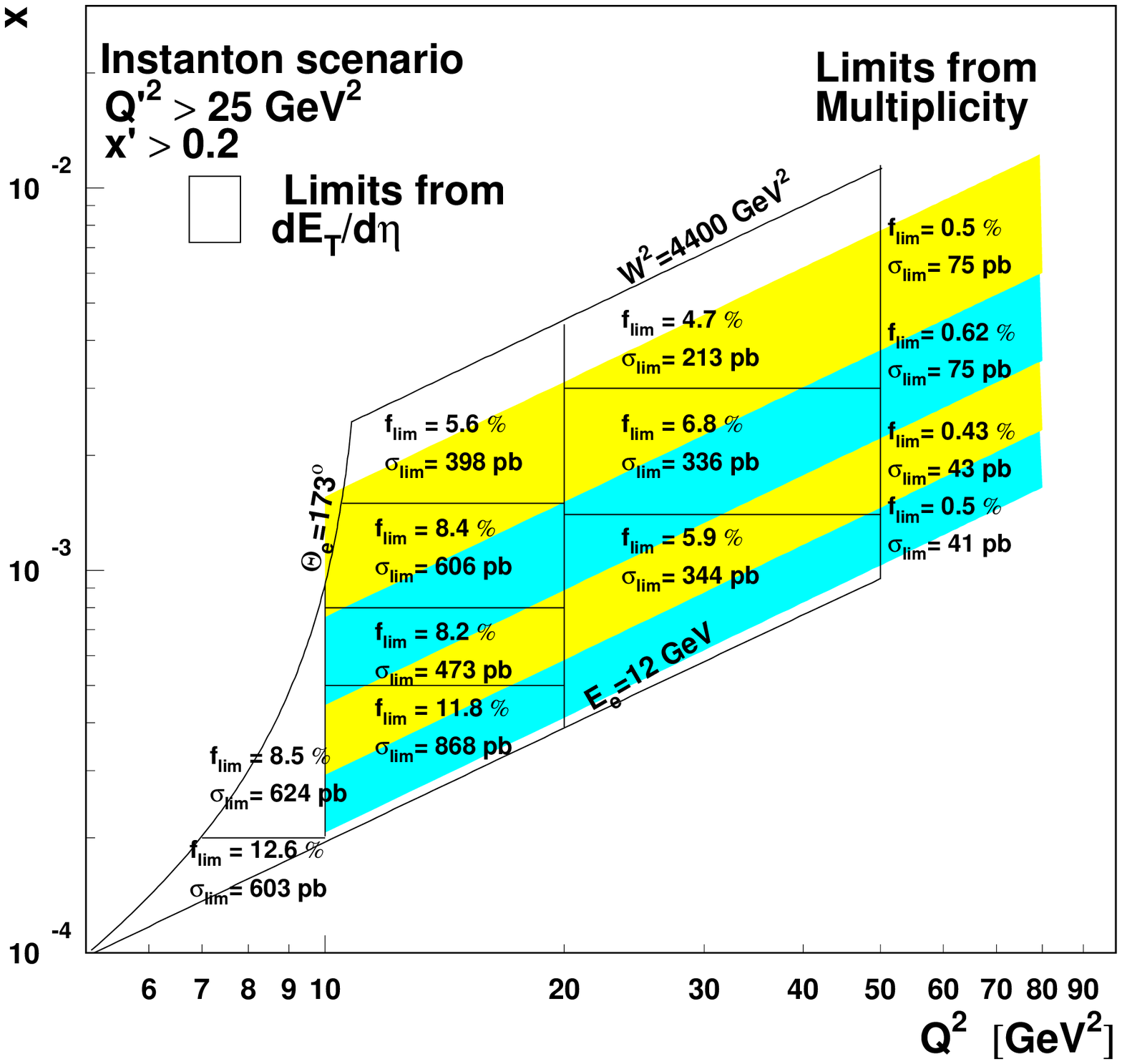,%
   width=12cm,bbllx=28pt,bblly=142pt,bburx=561pt,bbury=668,clip=}
 \fcaption{
      95\% C.L. limits on instanton production with $\qprimesq > 25\GeVsq$ and
      $\xprime > 0.2$ \cite{inst:bounds}.
      The cross-section limits ($\sigma_{\rm lim}$)
      together with the maximally allowed instanton fraction
      $f_{\rm lim}$ are shown in the \xb--\Qsq plane.
      They are obtained from the \et flow analysis
      (open fields),
      and from the multiplicity analysis (shaded fields) with their
      numbers at the right edge.
      The boundaries implied by the analysis cuts of the energy flow
      analysis in the angle and energy of the scattered electron,
      $\theta_e < 173^\circ$
      and $E_e>12\GeV$,
      and by the requirement
      $W^2>4400 \GeVsq$ are indicated.}
 \label{limplane}
\end{figure}

The fact that H1\cite{h1:mult} did not observe any events above a certain
multiplicity $\nmax$ has been exploited\cite{inst:bounds}
to place more stringent
limits on instanton production\footnote{The previous limits from the
H1 multiplicity analysis\cite{h1:mult}
were derived from the shape of the multiplicity
distribution for $n<\nmax$.} .
A significant fraction of instanton induced events would have
multiplicities $n>\nmax$ (compare fig.~\ref{dpdn}a).
Instanton fractions $\finst>0.4-0.6\%$
can therefore be excluded for $x'>0.2$
(see figs.~\ref{dpdn}b,~\ref{limplane}), and somewhat
lower \finst values for a lower cut-off
\mbox{$x'>0.1$ \cite{inst:bounds}.}
This search method has the advantage that, in contrast to the
previous shape comparisons, it does not rely on assumptions for standard
QCD event topologies, since no background needs to be subtracted.
Unavoidable of course is the dependence on the expected
instanton event shape, which may be even more uncertain than
the standard QCD event shapes.

\begin{table}[t]
\tcaption{Limits on QCD instantons in DIS. A fraction $\finst>\flim$
 of instanton induced events in DIS is excluded at 95\% C.L..}
\label{tab:instlim}
\small
\begin{tabular}{||c|c|c|c|c|c|c||}\hline\hline
analysis & \multicolumn{3}{c|}{DIS kinematics covered} &
           \multicolumn{2}{c|}{instanton scenario}     & limit \\
\hline
         & \Qsq (\GeVsqx) & $x$ & \W (GeV) & $Q'^2$(\GeVsqx) & $x'$ & \flim  \\
\hline
$K^0$ \cite{h1:k0} & 10 -- 70 & 0.001~ -- 0.01~ & 95 -- 230 &
               $\gtrsim 1$ &  $\gtrsim 0.2$ & 6 \% \\
multipl. \cite{h1:mult} & 10 -- 80 & 0.0007 -- 0.012 & 80 -- 115 &
                $\gtrsim 1$ &  $\gtrsim 0.2$ & 2.7 \% \\
\et flows \cite{inst:bounds} & ~5 -- 50 & 0.0001 -- 0.01~ & 65 -- 230 &
                   $>25$     & $>0.2$  & 5 -- 13 \% \\
multipl. \cite{inst:bounds} & 10 -- 80 & 0.0001 -- 0.01~ & 80 -- 220 &
                   $>25$     & $>0.2$  & 0.4 -- 0.6 \% \\
\hline\hline
\end{tabular}
\end{table}

The available bounds on instanton production are summarised in
tab.~\ref{tab:instlim}. The most stringent limits
for the theoretically ``safe'' scenario $x'>0.2$
are still a
factor 20 higher than what is predicted from the instanton theory,
see fig.~\ref{dpdn}b.
Limits for other scenarios can be found in\cite{inst:bounds}.
For $x'>0.1$ they are already below
the naive extrapolation into the
theoretically uncertain region, providing a constraint
for the theory and the holy grail function $F(x')$.

\section{Conclusion}                         Instanton transitions,
a yet unexplored facette of non-abelian gauge field theories, have been
discussed. While in the electroweak theory the $B+L$ violating effects
induced by instantons
are expected only at energies $\gtrsim 10\TeV$, their chirality violating
pendant in QCD could lead to striking signatures already at present day
colliders. In DIS at HERA, these are a high particle multiplicity
with large transverse energy localized in rapidity,
and $s$, $c$ and possibly $b$ quarks
in the final state.
The expected contribution to DIS events is of $\order{10^{-3}-10^{-4}}$,
with substantial theoretical uncertainties.
First analysis of HERA data taken in the years $\leq 1994$, corresponding
to an integrated luminosity of $\order{1.3\pbinv}$, are still a factor
$\approx 20$ above the prediction. With higher statistics data samples
($\order{25\pbinv}$ up to summer 1997)
and improved search strategies, a fundamental discovery at HERA
appears to be in reach.
It might be possible to exploit also other
reactions than DIS, such as photoproduction, where the hard scale needed for
reliable instanton calculations could be provided by the \pt of a jet.
\newpage

\renewcommand{\section}[1] {\vspace*{0.0cm}\addtocounter{sectionc}{1}
\setcounter{subsectionc}{0}\setcounter{subsubsectionc}{0}\noindent
        {\normalsize\bf\thesectionc. #1}\par\vspace*{0.0cm}}
\section{Acknowledgements}                   I would like to thank B. Kniehl and G. Kramer for their invitation
to this beautifully set workshop,
and I thank T. Carli, A. Ringwald and F. Schrempp
for the exciting time we are having with instantons, and for
their critical reading of the manuscript.

\renewcommand{\section}[1] {\vspace*{0.6cm}\addtocounter{sectionc}{1}
\setcounter{subsectionc}{0}\setcounter{subsubsectionc}{0}\noindent
        {\normalsize\bf\thesectionc. #1}\par\vspace*{0.4cm}}

\section{References}
\begin{footnotesize}

%
%
%
%
%

\end{footnotesize}


\end{document}